\documentclass[a4,11pt]{cip-submit}  
\usepackage{afterpage}
\usepackage{mathtools,slashed}

\usepackage{mathptmx}
\usepackage{hyperref}
\hypersetup{
colorlinks=false,
pdfborder={0 0 0}
}
\def\Mr{\uppercase}
\usepackage{float}
\usepackage{wasysym}
\usepackage{graphicx,wrapfig,tikz}
\usepackage{caption}
\usepackage{subcaption}
\usepackage{amsmath,amsxtra,
amssymb,latexsym,amscd,color,bm}
\usepackage[mathscr]{eucal}
\usepackage{epstopdf}
\usepackage{array}
\usepackage{cite}

\def\vsm{\vskip0.1cm}
\def\doi#1{\href{http://dx.doi.org/#1}{doi:#1}}

\def\titles#1{\title{\large\bf\noindent #1}}
\def\authors#1{\author{\begin{flushleft}{#1}\end{flushleft}}}
\def\authord#1#2{\indent\Mr{#1}\\
	\textit{\indent#2}\vsm}

\def\email#1{\bigskip\href{mailto:#1}{\textit{E-mail:}~{#1}}\\[3mm]}

\def\received#1{\vsm\textit{\indent Received #1}}
\def\accepted#1{\vsm\textit{Accepted for publication~#1}}

\def\and{$\text{\tiny AND }$}

\begin{document}
\titles{One-loop formulas for 
off-shell decay $H^* \rightarrow W^+W^-$
in 't Hooft-Veltman gauge and its applications}
\authors{\authord{Khiem Hong Phan, Dzung Tri Tran}
{
Institute of Fundamental and Applied Sciences, 
Duy Tan University, Ho Chi Minh City $700000$, Vietnam\\ 
Faculty of Natural Sciences, Duy Tan University, 
Da Nang City $550000$, Vietnam
}
\authord{Anh Thu Nguyen}
{
University of Science Ho Chi Minh City, $227$ 
Nguyen Van Cu, District $5$, Ho Chi Minh City, Vietnam
}
\email{phanhongkhiem@duytan.edu.vn}
\received{\today}
\accepted{DD MM YYYY}  
}
\maketitle
\markboth{Khiem Hong Phan }
{One-loop formulas for 
off-shell decay $H^* \rightarrow W^+W^-$
in 't Hooft-Veltman gauge and its applications}
\begin{abstract}
We present analytic results for 
one-loop radiative corrections to 
off-shell decay $H^* \rightarrow W^+W^-$ 
in 't Hooft-Veltman gauge within 
Standard Model framework.  
In numerical results, we show off-shell 
decay rate and its corrections with varying 
off-shell Higgs mass. The results show that 
the corrections are of $10\%$ contributions
to total decay rates. Furthermore, we study 
the impacts of one-loop radiative
corrections to off-shell decay 
$H^* \rightarrow W^+W^-$
in Higgs processes at future colliders. 
The signal processes such as
$e^-e^+\rightarrow
ZH^*\rightarrow Z(WW)$ with including 
the initial beam polarizations and 
$e^-e^+\rightarrow \nu_e\bar{\nu}_e H^*
\rightarrow \nu_e\bar{\nu}_e (WW)$ 
and $e^-\gamma \rightarrow e^-H^*
\rightarrow e^-WW$
are examined. We find that the 
effects are visible impacts
and these should be taken 
into account 
at future colliders. 
\end{abstract}
\section{Introduction}
One of the main targets of future 
colliders like the high-Luminosity 
Large Hadron Collider 
(HL-LHC)~\cite{Liss:2013hbb,
CMS:2013xfa}
and future 
colliders~\cite{Baer:2013cma} 
is to measure accurately the 
properties of standard-model-like 
Higgs boson. From the 
experimental data, we can verify 
the nature of the 
scalar Higgs potential and 
understand deeply the electroweak 
spontaneous symmetry breaking. 
It means that all Higgs 
productions and decay channels 
should be probed precisely as 
possible. 
Among Higgs decay processes, 
off-shell Higgs decay to 
W-boson pair are 
considerable interest at present 
and future
colliders~\cite{CDF:2010zen, 
CMS:2011egr,CMS:2012zbs,ATLAS:2011aa,
CMS:2013zmy,ATLAS:2015muc,ATLAS:2019vrd,
CMS:2022uhn, ATLAS:2022ooq, ATLAS:2022ndd}. 
Since the decay processes could
provide an important information
for understanding the Higgs 
sectors at higher-energy 
scales which are
sensitive with new 
physics contributions.

Full one-loop electroweak corrections to
Higgs decay to $W$-pair have calculated
in~\cite{Kniehl:1991xe} and to 
$H\rightarrow W^*W^*\rightarrow 4$
leptons have performed 
in~\cite{Ma:2021cxg,Kniehl:2012rz,
Bredenstein:2006rh,Bredenstein:2006ha}. 
The calculations for one-loop radiative 
corrections to 
$H\rightarrow W^*W^*\rightarrow 4$
leptons in many of extensions for the SM
have reported in~\cite{Hollik:2011xd, 
Altenkamp:2017kxk, Altenkamp:2017ldc, Altenkamp:2018bcs}. 
Due to the importance of the decay channel, 
we perform the calculations for one-loop 
radiative 
corrections to $H\rightarrow WW$ with 
the following extensions. 
We first provide analytic results for 
one-loop radiative 
corrections to $H\rightarrow WW$ for both 
off-shell Higgs and $W$-pair.
As a result, our results
are valid for one-loop correction to the vertex
$HWW$ which 
can be taken into account in many 
relevant process calculations.
Moreover, 
we can apply double-pole
approximation for studying off-shell Higgs decay 
to W-pair from the analytical results in this paper. 
In further, the study for the 
impacts of one-loop corrections to off-shell 
decay to $W$-pair through Higgs signal 
processes at future colliders like 
$e^-e^+$ collisions, photon-electron collisions.
In the detail, our computations are performed 
in 't Hooft-Veltman gauge in the framework of
the SM. One-loop form factors 
for the off-shell decay are expressed in terms 
of scalar one-loop Passarino-Veltman functions
in the standard notations of {\tt LoopTools}. 
As a result, off-shell decay rates can be 
evaluated numerically by using this program. 
In phenomenological results, we show decay 
rates and the corrections as a function of
off-shell Higgs mass. The results show that 
the corrections are of $10\%$ contributions. 
In addition, we study the 
impacts of one-loop corrections to 
off-shell decay $H^* \rightarrow W^+W^-$
in Higgs processes at future colliders. 
Specially, all signal 
processes such as
$e^-e^+\rightarrow
ZH^*\rightarrow Z(WW)$,
$e^-e^+\rightarrow f\bar{f}H^*
\rightarrow f\bar{f} (WW)$ for $f=e, \nu_e$
and $e^-\gamma \rightarrow e^-H^*
\rightarrow e^-WW$
are examined. In this analysis, we 
include the initial beam polarization effects
and consider both the unpolarized case 
and the case of longitudinal polarizations 
for $W$ bosons. 
We find that the effects are visible impacts
and these must be taken into account 
at future lepton colliders.

The layout of the paper is as follows: 
In section $2$, we first present 
one-loop expressions for the 
vertex $HWW$. 
We then apply these formulas
to the off-shell Higgs decay 
channel $H^*\rightarrow WW$. 
We also take into account for
soft and hard photon contributions
in this section. 
In section $3$, phenomenological 
results are shown. The impacts of 
one-loop 
off-shell Higgs decay through 
Higgs productions
at future colliders 
are discussed in this section. 
Conclusions
and outlook for this research 
are discussed in the section $4$. 
In the appendix, we show 
the one-loop 
counter-term for $HWW$ vertex. 
One-loop Feynman diagrams in 
't Hooft-Veltman gauge for this 
decay channel are 
shown in the appendix $C$.
\section{Calculations} 
We present in detail the calculations
in this section. All one-loop 
Feynman diagrams contributing to 
the vertex $HW^+W^-$ in 't Hooft-Veltman 
gauge can be classified into two 
groups (shown in appendix $C$).
First, we consider all fermions
propagating in the loop diagrams
to group $1$ (called as $G_1$). 
In the group $2$ (noted as $G_2$)
all $W$, $Z$ bosons, scalar $H$ boson,
goldstone bosons and ghost particles 
exchanging in the loop diagrams. 
As we shown later that one-loop
contributing to the vertex $HWW$ 
contains both ultraviolet divergent 
($UV$-divergent) and infrared divergent
(IR-divergent). The counter-terms 
for cancelling the $UV$-divergent 
are shown as diagram in group $G_0$. 
To
handle with the 
IR-divergent, we include the 
bremsstrahlung processes
$H^* \rightarrow W^+W^-\gamma$
for both soft and hard photon 
contributions.

In general, one-loop 
contributions for the 
$H(p)W^+(q_1)W^-(q_2)$ vertex
are
expressed in terms of the Lorentz
structure as follows:
\begin{eqnarray}
 \mathcal{V}_{HW^+W^-}^{1-\text{loop}}
 &=& g_{HWW} \Big\{ F_{00}\; g^{\mu\nu} 
 +\sum\limits_{i,j = 1}^2 
 F_{ij}\; q_i^{\nu}q_j^{\mu}
+ i F \varepsilon_{\mu\nu\rho\sigma} 
 q_1^{\rho}q_2^{\sigma}
 \Big\}.
\end{eqnarray}
Where $g_{HWW} = e M_W/s_W$ is the coupling
of Higgs to $W$-pair. Here
$W$ boson mass is $M_W$ and $c_W$ ($s_W$)
is cosine (sine)
of Weinberg angle, respectively. 
In the vertex, 
the scalar functions $F_{00}, F_{ij}$ 
for $i,j=1,2$ and $F$ are so-called 
one-loop form factors. 
They are functions of the momenta-squared 
as $p^2, q_1^2, q_2^2$. 
In this calculation, we use the 
{\tt Package-X}~\cite{Patel:2015tea} for 
handling all Dirac 
traces and Lorentz contractions in $d$ 
dimensions. One-loop amplitudes 
are then decomposed into tensor 
one-loop integrals which are 
expressed in terms of 
the scalar PV-functions~\cite{Denner:2005nn}
in standard notations
of {\tt LoopTools}~\cite{Hahn:1998yk}. 
As a result, one-loop form factors
can be evaluated numerically
by using this package.

In detail,
analytical results for all form factors
are shown in the following paragraphs. 
They are calculated as follows:
\begin{eqnarray}
F_{00} &=& \sum\limits_{G=\{G_0, 
G_1, G_2\}}F_{00}^{(G)}
\end{eqnarray}
with $\{G_0, G_1, G_2\} = 
\{\text{group 0}, \text{group 1}, 
\text{group 2}\}$
of Feynman diagrams. By considering 
the contributions 
of Feynman diagram in group $1$, 
we have 
\begin{eqnarray}
F_{00}^{(G_1)} &=& 
\dfrac{e^3}{(64 \pi^2) 
s_W^3 M_W} N^C_t m_t^2
\Bigg\{
2 B_0(p^2,m_t^2,m_t^2)
+B_0(q_1^2,m_b^2,m_t^2)
+B_0(q_2^2,m_b^2,m_t^2)
\\
&&\hspace{0cm}
+ \Big[ (2 m_t^2 + m_b^2 - q_1^2 - q_2^2) 
C_0(p^2,q_1^2,q_2^2,m_t^2,m_t^2,m_b^2)
-8 C_{00}(p^2,q_1^2,q_2^2,m_t^2,m_t^2,m_b^2)
\Big]
\Bigg\}.
\nonumber
\end{eqnarray}
It is noted that we take top and bottom
quarks in the loop diagrams as an example. 
Our results must be included all fermions
contributing in one-loop diagrams. 
From the second group of Feynman diagrams,
one arrives at 
\begin{eqnarray}
F_{00}^{(G_2)} &=& 
\dfrac{e^3}{(128 \pi^2) 
s_W^3 c_W^4 M_W}
\Bigg\{
\Big[
8 M_W^2 c_W^6 (4 d - 7) 
+ 4 c_W^4 (M_H^2 + 2 M_W^2 s_W^2)
\Big]
\nonumber\\
&&
\hspace{8.3cm}
\times 
C_{00}(p^2,q_1^2,q_2^2,M_W^2,M_W^2,M_W^2)
\nonumber\\
&&
+4 M_W^2 c_W^2 
\Big[
2 c_W^4 (M_W^2+M_Z^2)
-2 M_W^2 s_W^2 c_W^2
-M_H^2 s_W^4
\nonumber \\
&&
+ c_W^2 \Big((4 c_W^2+s_W^2) (q_1^2+q_2^2)
- p^2 (5 c_W^2+2 s_W^2)\Big)
\Big] 
C_0(p^2,q_1^2,q_2^2,M_W^2,M_W^2,M_W^2)
\nonumber\\
&&
+ 4 c_W^2 
\Big[
\Big(
M_H^2
+4 M_W^2 (2 d-3) - 2 M_W^2
\Big) c_W^2
+2 M_W^2 s_W^2
\Big] 
C_{00}(p^2,q_1^2,q_2^2,M_W^2,M_W^2,M_W^2)
\nonumber \\
&&
+4 M_W^2 
\Big[
2 c_W^4 M_Z^2
+\Big(4 (q_1^2+q_2^2)-5 p^2\Big) c_W^4
\Big] 
C_0(p^2,q_1^2,q_2^2,M_W^2,M_W^2,M_W^2)
\nonumber \\
&&
+ 32 (d - 2)M_W^2 s_W^2 c_W^4 
C_{00}(p^2,q_1^2,q_2^2,M_W^2,M_W^2,0) 
\nonumber\\
&&
+ 4 M_W^2 s_W^2 c_W^4\Big[(2 M_W^2-M_H^2)
+3 (q_1^2+q_2^2-p^2)\Big]
C_0(p^2,q_1^2,q_2^2,M_W^2,M_W^2,0) 
\nonumber \\
&&
+ 4 c_W^4 
\Big(M_H^2 + 2 M_W^2\Big)
C_{00}(p^2,q_1^2,q_2^2,M_W^2,M_W^2,M_H^2) 
- 8 c_W^4M_W^4 
C_0 (p^2,q_1^2,q_2^2,M_W^2,M_W^2,M_H^2)
\nonumber\\
&&
+ 12 M_H^2 c_W^4 
C_{00} (p^2,q_1^2,q_2^2,M_H^2,M_H^2,M_W^2)
- 12 M_W^2 M_H^2 c_W^4 
C_0(p^2,q_1^2,q_2^2,M_H^2,M_H^2,M_W^2)
\nonumber \\
&&
-3 M_H^2 c_W^4 B_0(p^2,M_H^2,M_H^2)
-4 M_W^2 c_W^4 
\Big[B_0(q_1^2,M_H^2,M_W^2)
+ B_0(q_2^2,M_H^2,M_W^2)
\Big]
\nonumber\\
&&
+4 M_W^2 c_W^2 \Big(c_W^4+c_W^2-s_W^4\Big) 
\Big[
B_0(q_1^2,M_W^2,M_W^2)
+ 
B_0(q_2^2,M_W^2,M_W^2)
\Big] 
\nonumber\\
&&
-\Big[
3M_H^2 c_W^4 -8M_W^2 c_W^4 (d-2)
\Big]
B_0(p^2,M_W^2,M_W^2)
\Bigg\}
. \nonumber
\end{eqnarray}
As we show in later, form factor
$F_{00}$ contains the UV-divergent.
Following renormalization theory, 
the counter-term ($F_{00}^{(G_0)}$) 
is given in Eq.~(\ref{counter}). 
Other form factors can be given 
as follows:
\begin{eqnarray}
 F_{ij} &=& 
\sum\limits_{G=\{G_1, G_2\}}F_{ij}^{(G)} 
\end{eqnarray}
Applying the same procedure, one has 
analytic expression for $F_{11}$ as
\begin{eqnarray}
F_{11}^{(G_1)} &=& 
-\dfrac{e^3}{(32 \pi^2) 
s_W^3 M_W} N^C_t m_t^2
\Big[
C_0(p^2,q_1^2,q_2^2,m_t^2,m_t^2,m_b^2)
+5 C_1(p^2,q_1^2,q_2^2,m_t^2,m_t^2,m_b^2)
\nonumber\\
&&
\hspace{4cm}
+4 C_{11}(p^2,q_1^2,q_2^2,m_t^2,m_t^2,m_b^2)
\Big]
\end{eqnarray}
and 
\begin{eqnarray}
F_{11}^{(G_2)} &=& 
\dfrac{e^3}{(32 \pi^2) s_W^3 c_W^2 M_W}
\Bigg\{
M_W^2 c_W^2 (2 c_W^2 + s_W^2) C_0(p^2,q_1^2,q_2^2,M_W^2,M_W^2,M_W^2)
\\
&&
+c_W^2 \Big[4 M_W^2 c_W^2 (2 d - 3) + M_H^2\Big] 
C_1(p^2,q_1^2,q_2^2,M_W^2,M_W^2,M_W^2)
\nonumber\\
&&
+ c_W^2 \Big[
2 M_W^2 c_W^2 (4 d - 7) +M_H^2 +2 M_W^2 s_W^2 
\Big] 
C_{11}(p^2,q_1^2,q_2^2,M_W^2,M_W^2,M_W^2 )
\nonumber\\
&&
+ 
2 M_W^2 C_0(p^2,q_1^2,q_2^2,M_W^2,M_W^2,M_W^2 )
\nonumber\\
&&
+\Big[
M_H^2 c_W^2
+ 4 M_W^2 c_W^2 (2 d - 3)
+3 M_W^2 s_W^2
\Big] 
C_1(p^2,q_1^2,q_2^2,M_W^2,M_W^2,M_W^2 )
\nonumber\\
&&
+ \Big[
M_H^2 c_W^2 
+ 2 M_W^2 c_W^2 (4 d -7) 
+2 M_W^2 s_W^2 
\Big] 
C_{11}(p^2,q_1^2,q_2^2,M_W^2,M_W^2,M_W^2 )
\nonumber\\
&&
+ 
2 c_W^2 M_W^2 
C_0(p^2,q_1^2,q_2^2,M_W^2,M_W^2,M_H^2 )
+c_W^2\Big(M_H^2 + 3 M_W^2\Big) C_1(p^2,q_1^2,q_2^2,M_W^2,M_W^2,M_H^2 )
\nonumber\\
&&
+c_W^2\Big(M_H^2 + 2 M_W^2\Big) C_{11}(p^2,q_1^2,q_2^2,M_W^2,M_W^2,M_H^2 )
\nonumber\\
&&
+3 M_H^2 c_W^2 \Big[
C_1(p^2,q_1^2,q_2^2,M_H^2,M_H^2,M_W^2 )
+ C_{11}
(p^2,q_1^2,q_2^2,M_H^2,M_H^2,M_W^2 ) 
\Big]
\nonumber\\
&&
+M_W^2 s_W^2 c_W^2 
\Big[
C_0(p^2,q_1^2,q_2^2,M_W^2,M_W^2,0 )
+(8 d - 12) 
C_1(p^2,q_1^2,q_2^2,M_W^2,M_W^2,0 )
\nonumber\\
&&\hspace{6cm}
+(8 d - 16) 
C_{11} (p^2,q_1^2,q_2^2,M_W^2,M_W^2,0 )
\Big]
\Bigg\}
. 
\nonumber
\end{eqnarray}
Analytical formulas for $F_{22}$ 
are shown accordingly
\begin{eqnarray}
F_{22}^{(G_1)} &=& 
-\dfrac{e^3 N^C_t m_t^2}{(32 \pi^2) s_W^3 M_W} 
\Big[
C_1 (p^2,q_2^2,q_1^2,m_t^2,m_t^2,m_b^2)
+4 C_{11} (p^2,q_2^2,q_1^2,m_t^2,m_t^2,m_b^2)
\Big]
, \\
F_{22}^{(G_2)} &=& 
\dfrac{e^3}{(32 \pi^2) s_W^3 c_W^2 M_W}
\Bigg\{
3 M_H^2 c_W^2 
C_{11} (p^2,q_2^2,q_1^2,M_H^2,M_H^2,M_W^2)
\\
&&
+ 
M_W^2 s_W^2 c_W^2 C_0(p^2,q_2^2,q_1^2,M_W^2,M_W^2,M_W^2) 
+2 M_W^2 c_W^4  C_1(p^2,q_2^2,q_1^2,M_W^2,M_W^2,M_W^2) 
\nonumber\\
&&
+ c_W^2\Big[
2 c_W^2 M_W^2 (4 d - 7) 
+ (M_H^2 + 2 M_W^2 s_W^2) 
\Big] C_{11} 
(p^2,q_2^2,q_1^2,M_W^2,M_W^2,M_W^2)
\nonumber\\
&&
+ M_W^2 (2 c_W^2 + s_W^2) 
C_1 (p^2,q_2^2,q_1^2,M_W^2,M_W^2,M_W^2)
\nonumber\\
&&
+ \Big[
c_W^2  \Big(2 M_W^2 (4 d -7)+M_H^2\Big)
+2 M_W^2 s_W^2
\Big] 
C_{11}(p^2,q_2^2,q_1^2,M_W^2,M_W^2,M_W^2)
\nonumber\\
&&
+ c_W^2 \Big[
M_W^2 C_1 (p^2,q_2^2,q_1^2,M_W^2,M_W^2,M_H^2)
+ (M_H^2 + 2 M_W^2) 
C_{11} (p^2,q_2^2,q_1^2,M_W^2,M_W^2,M_H^2)
\Big]
\nonumber\\
&&
+ M_W^2 s_W^2 c_W^2 \Big[
8(d - 2) 
C_{11}(p^2,q_2^2,q_1^2,M_W^2,M_W^2,0)
+ 2 C_1(p^2,q_2^2,q_1^2,M_W^2,M_W^2,0)
\nonumber\\
&&
- C_0 (p^2,q_2^2,q_1^2,M_W^2,M_W^2,0)  
\Big]
\Bigg\}. \nonumber
\end{eqnarray}

Form factor $F_{12}$ is given by
\begin{eqnarray}
F_{12}^{(G_1)} &=& 
\dfrac{e^3}{(8 \pi^2) s_W^3 M_W} N^C_t m_t^2
\Big[
C_1(p^2,q_2^2,q_1^2,m_t^2,m_t^2,m_b^2)
+C_{12}(q_1^2,p^2,q_2^2,m_b^2,m_t^2,m_t^2)
\Big]
\end{eqnarray}
and 
\begin{eqnarray}
F_{12}^{(G_2)} &=& 
\dfrac{e^3}{(32 \pi^2) s_W^3 c_W^2 M_W}
\Bigg\{
c_W^2 \Big[
2 M_W^2 c_W^2 (7-4d) 
- (M_H^2 + 6 M_W^2 s_W^2)
\Big]
\times \\
&& 
\hspace{6cm}
\times 
C_1 (p^2,q_2^2,q_1^2,M_W^2,M_W^2,M_W^2)
\nonumber\\
&&
\nonumber\\
&&
- 2 M_W^2 s_W^2 c_W^2 
\Big[
2 C_0(p^2,q_1^2,q_2^2,M_W^2,M_W^2,M_W^2)
+ 3 C_1(p^2,q_1^2,q_2^2,M_W^2,M_W^2,M_W^2)
\Big]
\nonumber\\
&&
+\Big[
2 M_W^2 c_W^2 (7-4d)
- (M_H^2 c_W^2     
+M_W^2 s_W^2) 
\Big]
C_1 (p^2,q_2^2,q_1^2,M_W^2,M_W^2,M_W^2)
\nonumber\\
&&
+M_W^2 s_W^2 
\Big[
C_0(p^2,q_1^2,q_2^2,M_W^2,M_W^2,M_W^2)
+ C_1 (p^2,q_1^2,q_2^2,M_W^2,M_W^2,M_W^2)
\Big]
\nonumber\\
&&
+\Big[
2 M_W^2 c_W^2 (7-4d)
- (M_H^2 c_W^2  
+ 2 M_W^2 s_W^2)
\Big]
C_{12}(q_1^2,p^2,q_2^2,M_W^2,M_W^2,M_W^2)
\nonumber\\
&&
+2 M_W^2 s_W^2 c_W^2 
\Big[
2C_0 (p^2,q_1^2,q_2^2,M_W^2,M_W^2,0)
+3 C_1 (p^2,q_1^2,q_2^2,M_W^2,M_W^2,0)
\nonumber\\
&&
- (4d - 10) 
C_1 (p^2,q_2^2,q_1^2,M_W^2,M_W^2,0)
\Big]
\nonumber\\
&&
+M_W^2 c_W^2 
\Big[
C_0 (p^2,q_1^2,q_2^2,M_W^2,M_W^2,M_H^2)
+C_1(p^2,q_1^2,q_2^2,M_W^2,M_W^2,M_H^2)
\nonumber\\
&&
- 8 s_W^2(d-2)
C_{12} (q_1^2,p^2,q_2^2,0,M_W^2,M_W^2)
\Big]
\nonumber\\
&&
- c_W^2 
\Big[
(M_H^2 
+M_W^2)
C_1 (p^2,q_2^2,q_1^2,M_W^2,M_W^2,M_H^2)
\nonumber\\
&& \hspace{4cm}
+ (M_H^2  
+2 M_W^2 )
C_{12} (q_1^2,p^2,q_2^2,M_H^2,M_W^2,M_W^2)
\Big]
\nonumber\\
&&
-3 M_H^2 c_W^2
\Big[ C_1 (p^2,q_2^2,q_1^2,M_H^2,M_H^2,M_W^2)
+ C_{12} (q_1^2,p^2,q_2^2,M_W^2,M_H^2,M_H^2)
\Big]
\nonumber\\
&&
-
\Big[ M_H^2 c_W^2 
+ 2 M_W^2 c_W^2   
\Big((4 d-7) c_W^2+s_W^2\Big)
\Big]
C_{12} (q_1^2,p^2,q_2^2,M_W^2,M_W^2,M_W^2)
\Bigg\}
. \nonumber
\end{eqnarray}
Next form factor $F_{21}$ is shown
\begin{eqnarray}
F_{21}^{(G_1)} &=& 
\dfrac{e^3\; N^C_t m_t^2}{(32 \pi^2) s_W^3 M_W} 
\Big[
C_1(p^2,q_1^2,q_2^2,m_t^2,m_t^2,m_b^2)
+C_1(p^2,q_2^2,q_1^2,m_t^2,m_t^2,m_b^2)
\nonumber\\
&&
\hspace{3cm}
+4 C_{12}(q_1^2,p^2,q_2^2,m_b^2,m_t^2,m_t^2)
\Big].
\end{eqnarray}
\begin{eqnarray}
F_{21}^{(G_2)} &=& 
\dfrac{e^3}{(32 \pi^2) s_W^3 c_W^2 M_W}
\Bigg\{ 
2 M_W^2 c_W^2 \Big[
4 C_0 (p^2,q_1^2,q_2^2,M_W^2,M_W^2,M_W^2)
\nonumber\\
&&
\hspace{6cm}
+(3 s_W^2 - c_W^2) C_1(p^2,q_1^2,q_2^2,M_W^2,M_W^2,M_W^2)
\Big]
\nonumber\\
&&
+2 M_W^2 c_W^2 \Big(3 s_W^2 - c_W^2 \Big)
C_1 (p^2,q_2^2,q_1^2,M_W^2,M_W^2,M_W^2)
\nonumber\\
&&
+\Big[
2 M_W^2 c_W^2 (7 - 4 d) 
- (M_H^2 c_W^2  
+2 M_W^2 s_W^2) 
\Big]
C_{12} 
(q_1^2,p^2,q_2^2,M_W^2,M_W^2,M_W^2)
\nonumber\\
&&
+2 M_W^2 \Big[
4 c_W^2 C_0(p^2,q_1^2,q_2^2,M_W^2,M_W^2,M_W^2)
-C_1 (p^2,q_1^2,q_2^2,M_W^2,M_W^2,M_W^2)
\Big]
\nonumber\\
&&
-2 M_W^2 
C_1 (p^2,q_2^2,q_1^2,M_W^2,M_W^2,M_W^2)
+8 M_W^2 s_W^2 c_W^2 (2 - d) C_{12} (q_1^2,p^2,q_2^2,0,M_W^2,M_W^2)
\nonumber\\
&&
-8 M_W^2 s_W^2 c_W^2 
\Big[
C_1 (p^2,q_1^2,q_2^2,M_W^2,M_W^2,0)
+ C_1 (p^2,q_2^2,q_1^2,M_W^2,M_W^2,0)
\Big]
\nonumber\\
&&
-2 M_W^2 c_W^2 
\Big[
C_1 (p^2,q_1^2,q_2^2,M_W^2,M_W^2,M_H^2)
+ C_1 (p^2,q_2^2,q_1^2,M_W^2,M_W^2,M_H^2)
\Big]
\nonumber\\
&&
-(M_H^2 + 2 M_W^2) c_W^2 C_{12} (q_1^2,p^2,q_2^2,M_H^2,M_W^2,M_W^2)
\nonumber\\
&& 
-3 M_H^2 c_W^2 C_{12} 
(q_1^2,p^2,q_2^2,M_W^2,M_H^2,M_H^2)
\nonumber\\
&&
- c_W^2 \Big[
M_H^2+2 M_W^2 
\Big(
c_W^2 (4 d-7) +s_W^2
\Big)
\Big] 
C_{12} (q_1^2,p^2,q_2^2,M_W^2,M_W^2,M_W^2)
\Bigg\}
. 
\end{eqnarray}
The form factor $F$ (the 
coefficient of Levi Civita) 
is written as
\begin{eqnarray}
F &=&
- \dfrac{e^3 N^C_t m_t^2}{(32 \pi^2) s_W^3 M_W 
\Big[p^4-2 p^2 
(q_1^2+q_2^2)+(q_1^2-q_2^2)^2\Big]} 
\times
\\
&&\times
\Bigg\{
\big(q_1^2-q_2^2\big) 
\Big[
\Big(2 m_t^2 + m_b^2 -p^2+q_1^2+q_2^2\Big) 
C_0(p^2,q_1^2,q_2^2,m_t^2,m_t^2,m_b^2)
-2 B_0(p^2,m_t^2,m_t^2)
\Big]
\nonumber\\
&&
+\Big(p^2-q_1^2-3 q_2^2\Big) 
B_0(q_2^2,m_b^2,m_t^2)
-\Big(p^2-3 q_1^2-q_2^2\Big) 
B_0(q_1^2,m_b^2,m_t^2)
\Bigg\}
. \nonumber
\end{eqnarray}
In the case of both external $W$-bosons 
are on-shell masses $q_1^2 = q_2^2 = M_W^2$, 
the form factor $F$ is canceled 
analytically in our calculations.

\subsection{One-loop virtual 
$H^*\rightarrow WW$}
We turn out attention to the one-loop 
amplitude for off-shell 
$H^* \rightarrow W^+W^-$. All one-loop 
virtual, soft and hard bremsstrahlung 
contributions are taken into account in
the current calculation. For one-loop 
virtual contributions, the amplitude 
is written in the form of
\begin{eqnarray}
 \mathcal{M}_{H^*\rightarrow 
 WW}^{1-\text{loop}}
 &=& g_{HWW} 
 \Big\{F_{00, H^*\rightarrow WW}\; g^{\mu\nu} 
 + F_{21, H^*\rightarrow WW}\; q_2^{\mu}q_1^{\nu}
 \Big\}\epsilon^*_{\mu}(q_1)\epsilon^*_{\nu}(q_2).
\end{eqnarray}
Where $\epsilon_{\mu},
\; \epsilon_{\nu}$ are
polarization vectors
for final $W$ bosons. 
Since one considers two real 
$W$ bosons in final state, 
we have only $F_{00}, F_{21}$ 
contributing to the amplitude. 
Analytic expressions
for these form factors 
can be written as follows:
\begin{eqnarray}
F_{00,H^*\rightarrow WW }  
&=& F_{00} (p^2; M_W^2,M_W^2 )
=\sum\limits_{G=\{G_{0}, 
G_1, G_2\}}F_{00}^{(G)}
(p^2; M_W^2, M_W^2),
\\
F_{21,H^*\rightarrow WW }  
&=& F_{21} (p^2; M_W^2,M_W^2 )
=\sum\limits_{G=\{ 
G_1, G_2\}}F_{21}^{(G)}
(p^2; M_W^2, M_W^2).
\end{eqnarray}
One-loop virtual off-shell decay rates 
for 
$H^* \rightarrow WW$ are calculated 
in terms of the above form factors 
as follows. Following kinematic 
variables $p^2 = M_{WW}^2, 
q_1^2 = M_W^2$ and $q_2^2 = M_W^2$
are used. For tree-level decay 
rates, we have
\begin{eqnarray}
 \Gamma_{\textrm{tree}}
&=&g_{HWW}^2 
\dfrac{
\sqrt{M_{WW}^2 -4M_W^2 }}
{64 \pi M_W^4 M_{WW}^2}
(12 M_W^4 - 4 M_W^2 M_{WW}^2
+ M_{WW}^4).
\end{eqnarray}
One-loop virtual decay rates 
are given by
\begin{eqnarray}
\Gamma_{\textrm{1-loop}}
&=&g_{HWW}^2 
\dfrac{
\sqrt{M_{WW}^2 -4M_W^2 }}
{64 \pi M_W^4 M_{WW}^2}
\Bigg\{ 
(24 M_W^4 - 8 M_W^2 M_{WW}^2  + 2 M_{WW}^4)
\mathcal{R}e [F_{00, H^*\rightarrow WW} ]
+
\\
&&
\hspace{4cm}
+ M_{WW}^2  (8 M_W^4 
- 6 M_W^2 M_{WW}^2  
+ M_{WW}^4)
\mathcal{R}e[
F_{21, H^*\rightarrow WW}
]
\Bigg\}.
\nonumber
\end{eqnarray}
\subsection{Soft photon  
contribution for 
$H^* \rightarrow W^+W^-\gamma_S$ }   
In order to regular IR-divergent, 
we have to include the soft contribution 
which is corresponding to the decay process  
$H^* (p) \rightarrow W^+_\mu (q_1) 
W^-_\nu (q_2) \gamma_\rho (q_3)$. 
The decay rates for the soft-photon 
contributions can be factorized 
as follows:
\begin{eqnarray}
\Gamma_{\text{soft}} = 
\delta_{\text{soft}} \Gamma_{\text{tree}}
\end{eqnarray}
where $\Gamma_{\text{tree}}$ is decay rates of
$H^* (p) \rightarrow W^+_\mu 
(q_1) W^-_\nu (q_2)$ at tree-level. 
The soft factor $\delta_s$ depends 
on the infrared regulator of photon mass 
$\lambda$ and photon energy cut-off
$k_c$ is given~\cite{Denner:2005nn}:
\begin{eqnarray}
\delta_{\text{soft}}
=
- \dfrac{\alpha}{4 \pi^2}
\int \limits_{\lambda \leq
E_\gamma \leq k_c}
\dfrac{d^3 q_3}{E_\gamma}
\Big(
\dfrac{q_1}{q_1 \cdot q_3}
-
\dfrac{q_2}{q_2 \cdot q_3}
\Big)^2
=
- \dfrac{\alpha}{4 \pi^2}
\Big[
I_{11}
+ I_{22}
- 2 I_{12}
\Big],
\end{eqnarray}
where photon energy 
$E_\gamma = \sqrt{|q_3|^2 
+ \lambda^2}$. The basic 
integrals $I_{ij}$ are given 
\begin{eqnarray}
I_{11}
&=& 
I_{22}
=
(2 \pi)
\Bigg\{
\ln \dfrac{4 k_c^2}{\lambda^2}
+
\dfrac{1}{\beta}
\ln \Big[
\dfrac{1-\beta}{1+\beta}
\Big]
\Bigg\}
,
\\
I_{12}
&=& 
\dfrac{(2 \pi)}{\beta}
\Bigg[
\beta^2 + \dfrac{2 M_W^2}{M_{WW}^2}
\Bigg]
\Bigg\{
\ln \Big[
\dfrac{1+\beta}{1-\beta}
\Big]
\ln \dfrac{4 k_c^2}{\lambda^2}
- 2 \text{Li}_2 \Big[
\dfrac{2\beta}{1+\beta}
\Big]
- \dfrac{1}{2}
\ln^2 \Big[
\dfrac{1+\beta}{1-\beta}
\Big]
\Bigg\}
\end{eqnarray}
where $\beta = \sqrt{1 - 4 M_W^2/M_{WW}^2}$. 
\subsection{Hard photon 
contribution for 
$H^* \rightarrow W^+W^-\gamma_H$}
\begin{figure}[ht]
\centering
\includegraphics[width=14.0cm, 
height=5cm]
{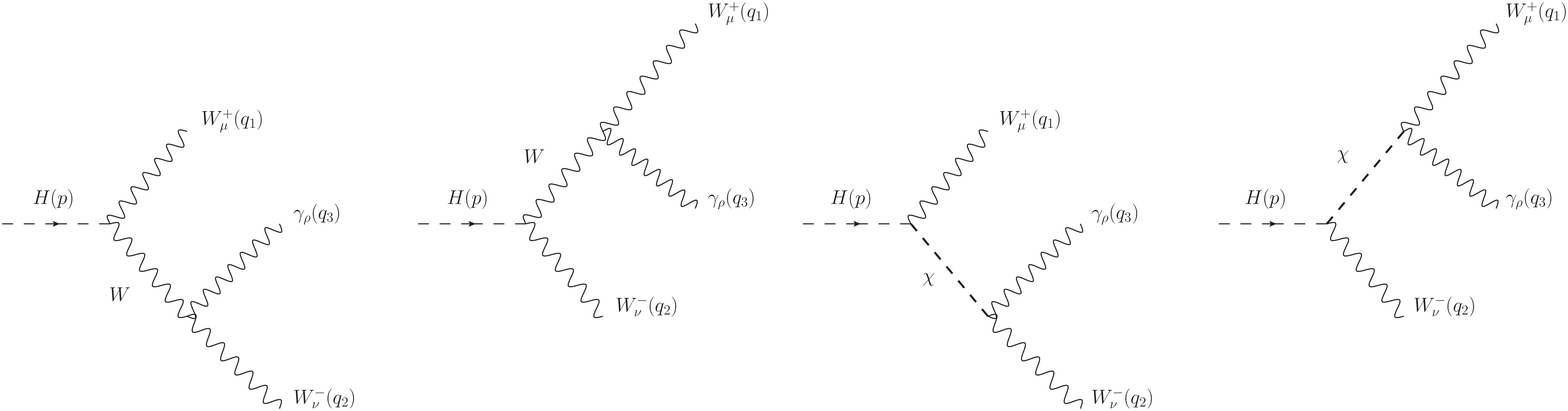}
\caption{\label{hardfeyn}
Tree-level Feynman
diagrams for hard photon contribution}
\end{figure}

We next consider hard-photon 
contribution in this subsection. 
The corresponding decay 
process is $H^* (p) \rightarrow 
W^+_\mu (q_1) W^-_\nu (q_2) \gamma_\rho (q_3)_H$ 
which $\gamma_H$ is hard photon
in final state. All tree-level Feynman 
diagrams
are plotted in Fig.~\ref{hardfeyn}.
The squared amplitude is shown in
terms of Mandelstam variables 
$s = (q_1 + q_2)^2, t = (q_2 + q_3)^2, 
u = (q_1 + q_3)^2$ as follows:
\begin{eqnarray}
\sum \limits_{\text{pol}}
|\mathcal{M_{\text{hard}}}|^2
&=&
\dfrac{e^4 M_W^2}
{M_W^2 (M_{WW}^2-M_W^2) 
(M_W^2-t)^2 (M_W^2-u)^2}
\times
\\
&&\hspace{0cm} \times
\Bigg\{
80 M_W^{10}-8 M_W^8 (7 s+12 (t+u))
+4 M_W^6 \Big[3 s^2 + 
18 s (t+u)+4 (2 t+u) (t+2 u)\Big]
\nonumber\\
&&\hspace{0cm}
-M_W^4 
\Big[s^3
+(14 s^2 + 16 t u) (t+u)
+s (25 t^2+74 t u+25 u^2)
\Big]
-s t u (s+t+u)^2
\nonumber\\
&&\hspace{0cm}
+M_W^2 \Big[s^3 (t+u)
+s (s+t+u) (3 t^2+14 t u+3 u^2)
-(t^2-u^2)^2\Big]
\Bigg\}.
\nonumber
\end{eqnarray}
The Mandelstam invariants follow
$s + t + u = M_{WW}^2 + 2 M_W^2$.
The decay rates are 
calculated accordingly
\begin{eqnarray}
\Gamma_{\text{hard}} = 
\dfrac{1}{256 \pi^3 M_{WW}^3 }
\int \limits_{4 M_W^2}^{M_{WW}(M_{WW}-2 k_c) } ds
\int \limits_{t_{\text{min}}}^{t_{\text{max}}} dt
\; 
\sum \limits_{\text{pol}}
|\mathcal{M_{\text{hard}}}|^2
,
\end{eqnarray}
where
\begin{eqnarray}
t_{\text{max},\text{min}}
=
\dfrac{1}{2}
\Bigg\{
M_{WW}^2 + 2 M_W^2 - s
\pm
\sqrt{\Big(1 - \dfrac{4 M_W^2}{s}\Big)
\Big[(M_{WW}^2 - s)^2 \Big]}
\Bigg\}
.
\end{eqnarray}
Having all the contributions, the total 
decay rate is given
\begin{eqnarray}
 \Gamma_{H^* \rightarrow 
 W^+W^-}^{\textrm{total}}=
 \Gamma_{\textrm{tree}} +
 \Gamma_{\textrm{1-loop}} 
 +
 \Gamma_{\textrm{soft}} + 
 \Gamma_{\textrm{hard}}. 
\end{eqnarray}
\section{Phenomenological results}         
We take the input parameters for 
phenomenological results as follows:
$M_Z = 91.1876$ GeV, 
$\Gamma_Z  = 2.4952$ GeV, 
$M_W = 80.379$ GeV, $\Gamma_W  = 2.085$ GeV, 
$M_H =125$ GeV, $\Gamma_H =4.07\cdot 10^{-3}$ GeV. 
The lepton masses are given: $m_e =0.00052$ GeV,
$m_{\mu}=0.10566$ GeV and $m_{\tau} = 1.77686$ GeV.
For quark masses, one takes $m_u= 0.00216$ GeV
$m_d= 0.0048$ GeV, $m_c=1.27$ GeV, $m_s = 0.93$ GeV, 
$m_t= 173.0$ GeV, and $m_b= 4.18$ GeV. 
Working in the so-called $G_{\mu}$-schemes, 
the Fermi constant 
is taken $G_{\mu}=1.16638\cdot 10^{-5}$ GeV$^{-2}$. 
The electroweak coupling can be then evaluated by 
\begin{eqnarray}
 \alpha = \sqrt{2}/\pi G_{\mu} M_W^2(1-M_W^2/M_Z^2)
 =1/132.184.
\end{eqnarray}
\subsection{Decay rates of 
off-shell $H^* \rightarrow W^+W^-$}
We first evaluate decay rates of 
off-shell $H^* \rightarrow W^+W^-$. 
In Fig.~\ref{decay500}, we present
decay rates of 
off-shell $H^* \rightarrow W^+W^-$
as a function of $M_{WW}$. Off-shell
Higgs mass $M_{WW}$ is range of 
$200$ GeV to $500$ GeV. In the left 
panel, tree-level contribution to
decay rates is presented as solid 
line. Total one-loop radiative 
corrections to the decay rates in 
the case of unpolarized for $W$ bosons 
is plotted as dashed-line and 
in the case of longitudinal polarizations
for $W$ bosons is shown
as dash-dotted line, respectively.
In the right panel, we show
the one-loop radiative corrections 
in percentage to the decay rates. 
The corrections are defined
as follows:
\begin{eqnarray}
 \delta[\%] =\dfrac{\Gamma_{H^*\rightarrow WW}^{\textrm{total}} 
 -\Gamma_{H^*\rightarrow WW}^{\textrm{tree}}
 }{\Gamma_{H^*\rightarrow WW}^{\textrm{tree}}}
 \times 100\%. 
\end{eqnarray}
\begin{figure}[ht]
\centering
$\begin{array}{cc}
\hspace{-4.2cm}
\Gamma_{H^*\rightarrow WW} [\text{GeV}] &
\hspace{-5.5cm}
\delta [\%]
\\
\includegraphics[width=7cm,height=6cm]
{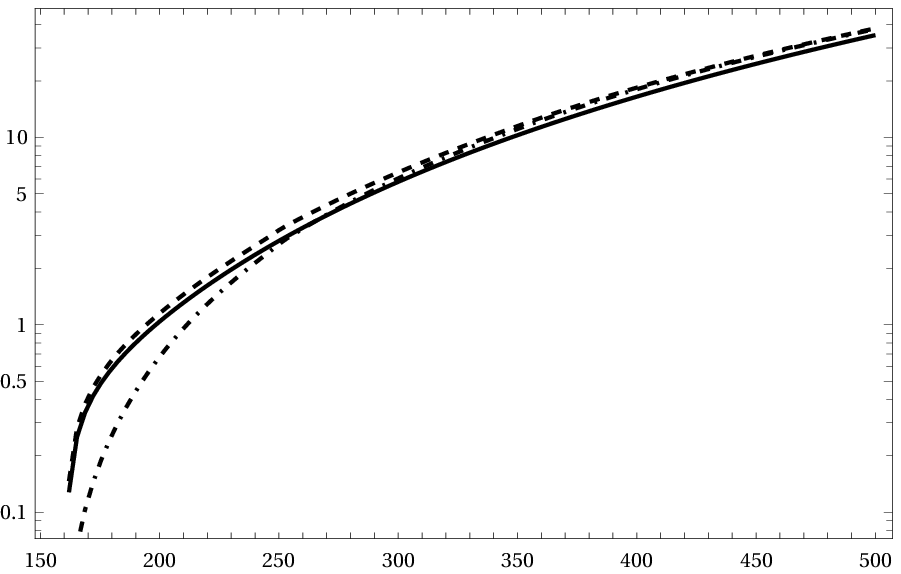}
& 
\includegraphics[width=7cm,height=6cm]
{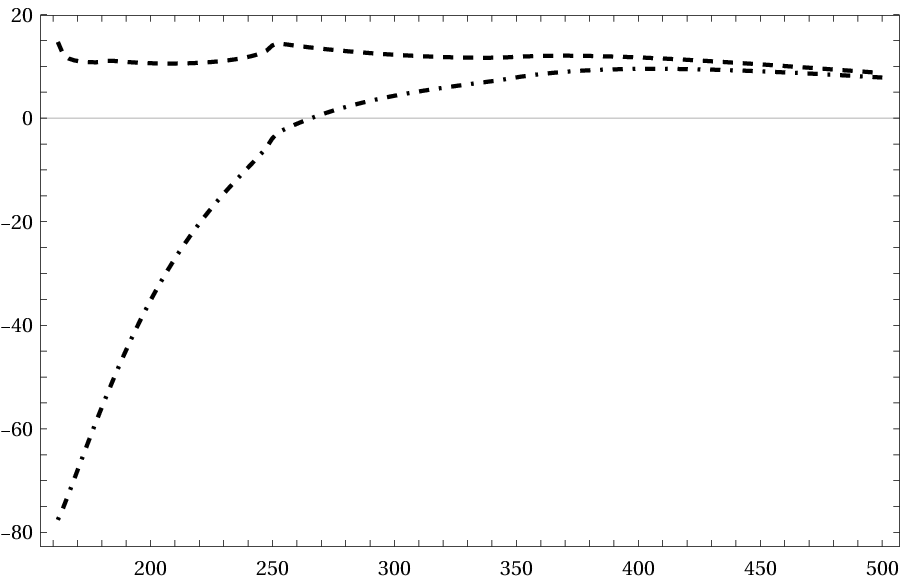}
\\
\hspace{5.2cm} M_{WW} [\text{GeV}]
&
\hspace{5.2cm} M_{WW} [\text{GeV}]
\end{array}$
\caption{\label{decay500} Off-shell Higgs 
decay rates as a function of 
$M_{WW}$.}
\end{figure}
We find that the corrections are range
form $5\%$ to $15\%$ for the unpolarized
case for $W$ bosons and from 
$-80\%$ to $10\%$ for the longitudinal
polarization
case for $W$ bosons, respectively. 
The corrections are massive 
contributions. They must be taken
into account at future colliders. 
\subsection{Off-shell                  
$H^* \rightarrow W^{+}W^{-}$           
in Higgs processes at future colliders}
\begin{figure}[H]
\centering
$\begin{array}{cc}
\hspace{-3.2cm}
d\sigma_{LR}/d M_{WW} [\text{fb/GeV}] 
&
\hspace{-3.2cm}
d\sigma_{LR}/d M_{WW} [\text{fb/GeV}] 
\\
\includegraphics[width=7cm,height=6cm]
{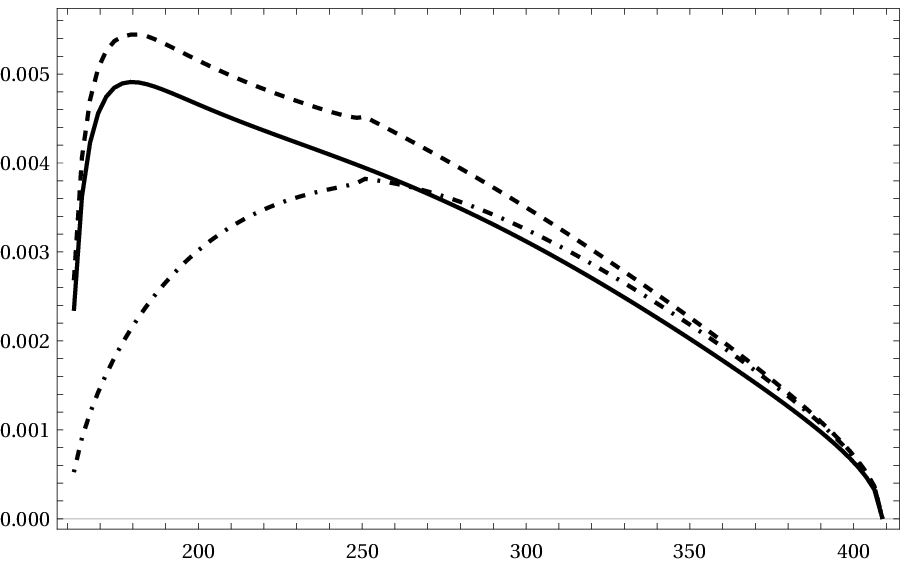}
& 
\includegraphics[width=7cm,height=6cm]
{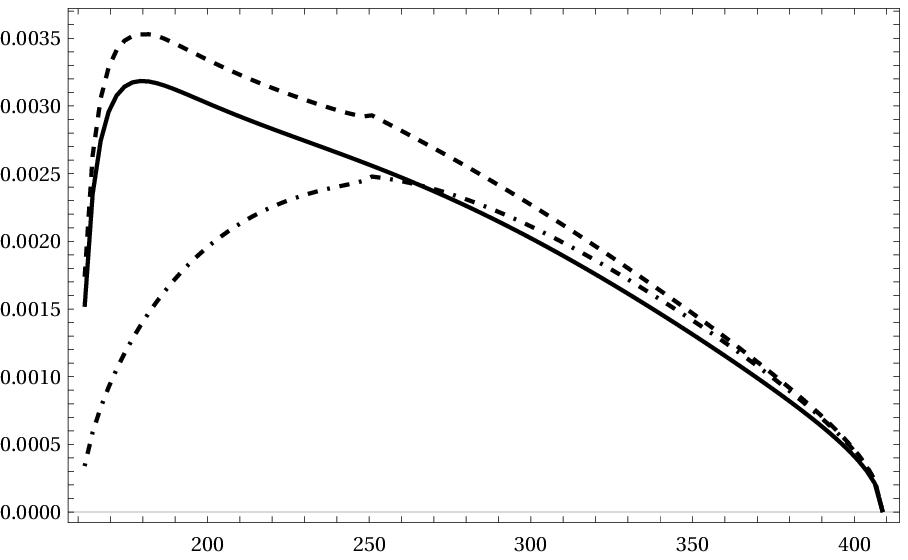}
\\
\hspace{5.2cm} M_{WW} [\text{GeV}]
&
\hspace{5.2cm} M_{WW} [\text{GeV}]
\\
&
\\
\hspace{-3.2cm}
d\sigma_{LR}/d M_{WW} [\text{fb/GeV}] 
&
\hspace{-3.2cm}
d\sigma_{LR}/d M_{WW} [\text{fb/GeV}] 
\\
\includegraphics[width=7cm,height=6cm]
{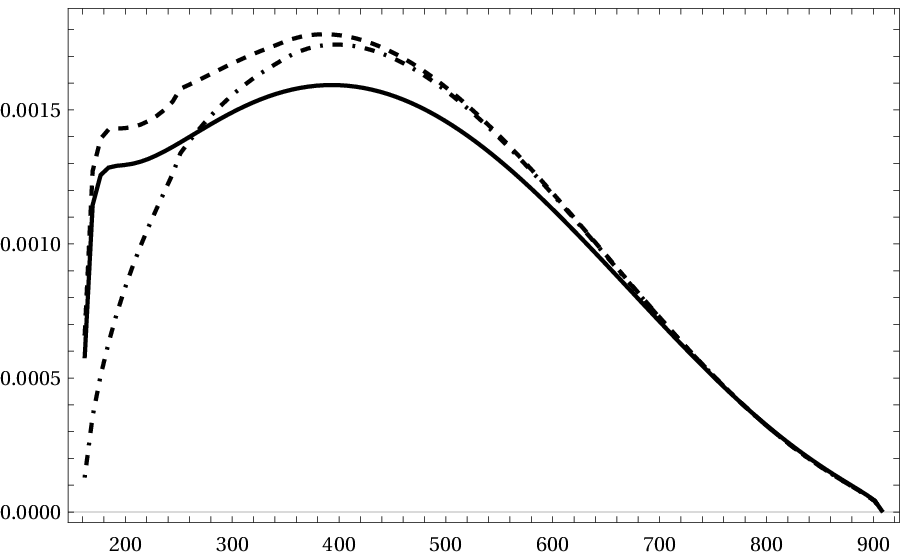}
& 
\includegraphics[width=7cm,height=6cm]
{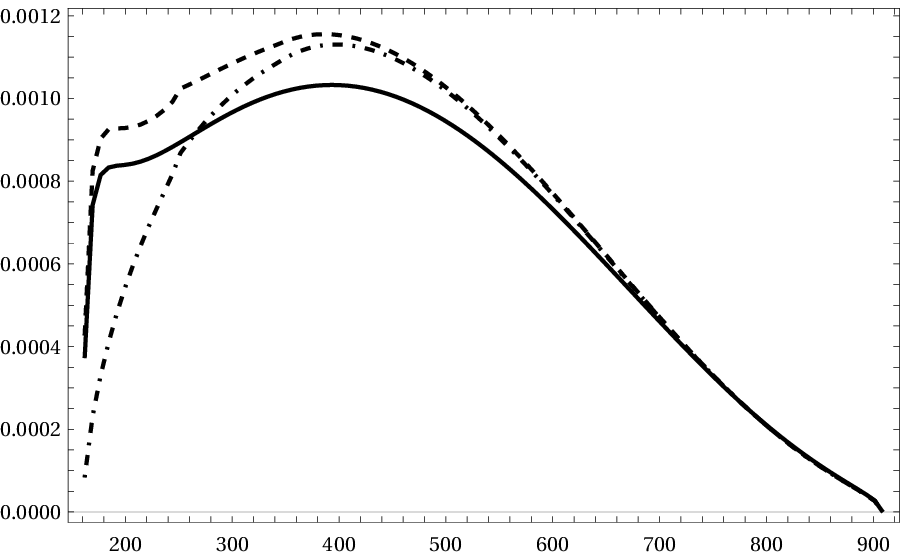}
\\
\hspace{5.2cm} M_{WW} [\text{GeV}]
&
\hspace{5.2cm} M_{WW} [\text{GeV}]
\end{array}$
\caption{\label{ILCZH} Differential 
cross sections as
a function of $M_{WW}$.}
\end{figure}

The effects of one-loop
off-shell $H^* \rightarrow 
W^{+}W^{-}$
in Higgs processes at future 
lepton colliders
are discussed in this section.
It is well-known that 
three main Higgs productions
at future lepton colliders, 
for example, 
$e^-e^+\rightarrow
ZH^*\rightarrow Z(WW)$ and 
$e^-e^+\rightarrow f\bar{f}H^*
\rightarrow f\bar{f} (WW)$ for 
$f=e, \nu_e$.
Since dominant cross sections
of $e^-e^+\rightarrow
ZH^*\rightarrow Z(WW)$ and 
$e^-e^+\rightarrow \nu_e\bar{\nu}_e H^*
\rightarrow \nu_e\bar{\nu}_e (WW)$
in comparision with
$e^-e^+\rightarrow e^-e^+ H^*
\rightarrow e^-e^+(WW)$ at the ILC,
we only shown numerical results 
for the signals of the former 
processes. 
All the signals are presented including
initial beam polarization at future 
lepton colliders. In detail,  
production cross sections 
are derived according to
\begin{eqnarray}
\label{masterformulas1}
\dfrac{d\sigma^{e^-e^+\rightarrow VH^*
\rightarrow V (WW)} (\sqrt{s})
}{dM_{WW}}
&=& 
\dfrac{2M_{WW}^2}{\pi}
\;
\dfrac{\sigma^{e^-e^+\rightarrow 
V H^*} (\sqrt{s}, M_{WW})  
\times 
\Gamma_{H^*\rightarrow WW} (M_{WW})
}
{[(M_{WW}^2-M_H^2)^2 + \Gamma_H^2 M_H^2 ]} 
\end{eqnarray}
for $V \equiv Z, f\bar{f}$ 
with $f\equiv e^-, \nu_e$. 
For deriving the above formulas, 
we refer our previous work
for more detail~\cite{Phan:2022amy, 
Nguyen:2022ubf}. 
Total cross sections then read:
\begin{eqnarray}
\sigma^{e^-e^+
\rightarrow  V(WW)}(\sqrt{s}) 
&=&\int\limits^{\sqrt{s}-M_V}_{2M_W} dM_{WW}\; 
\dfrac{2M_{WW}^2}{\pi}
\;
\dfrac{\sigma^{e^-e^+\rightarrow 
V H^*} (\sqrt{s}, M_{WW})  
\times 
\Gamma_{H^*\rightarrow WW} (M_{WW})
}
{[(M_{WW}^2-M_H^2)^2 + \Gamma_H^2 M_H^2 ]} .
\nonumber\\
\end{eqnarray}
In Fig.~\ref{ILCZH}, differential cross 
sections for the production 
$e^-e^+\rightarrow
ZH^*\rightarrow Z(WW)$
as function of off-shell Higgs
mass $M_{WW}$ are plotted at center-of-mass 
energy $\sqrt{s}=500$ GeV (two above Figures)
and at $\sqrt{s}=1000$ GeV (two below Figures). 
In the left (right) panel, we show for $LR$ ($RL$) 
polarization of $e^-, e^+$ beams, respectively. 
In these Figures, 
solid line is for tree-level contributions.
The dashed line presents for full one-loop
radiative corrections decay rates in the
case of unpolarized $W$-pair in final state. 
While the dash-dotted line shows for full one-loop
radiative corrections decay rates with longitudinal
polarization for $W$-pair.  Cross sections increase
up to the threshold $M_{WW} \sim 180$ GeV 
(for $\sqrt{s}=500$ GeV) and $M_{WW} \sim 400$ GeV 
(for $\sqrt{s}= 1000$ GeV), they decrease 
rapidly beyond 
the peaks. We find that one-loop corrections
to off-shell Higgs decay are visible impacts 
(specially, in the lower regions of off-shell Higgs mass)
in these distributions. 
In Fig.~\ref{ILCZH}, differential cross 
sections for the production 
$e^-e^+\rightarrow
\nu_e \bar{\nu}_e H^*\rightarrow \nu_e 
\bar{\nu}_e (WW)$
as function of off-shell Higgs
mass $M_{WW}$ are generated at 
$\sqrt{s}=500$ GeV (left panel) 
and at $\sqrt{s}= 1000$ GeV 
(right panel), respectively. 
We use same previous notations. 
We observe a peak around 
the threshold $M_{WW} = 2 M_W \sim 180$ GeV. 
Cross sections develop to the peak
and decrease rapidly beyond the peak.
\begin{figure}[ht]
\centering
$\begin{array}{cc}
\hspace{-3.2cm}
d\sigma_{LR}/d M_{WW} [\text{fb/GeV}] 
&
\hspace{-3.2cm}
d\sigma_{LR}/d M_{WW} [\text{fb/GeV}] 
\\
\includegraphics[width=7cm,height=6cm]
{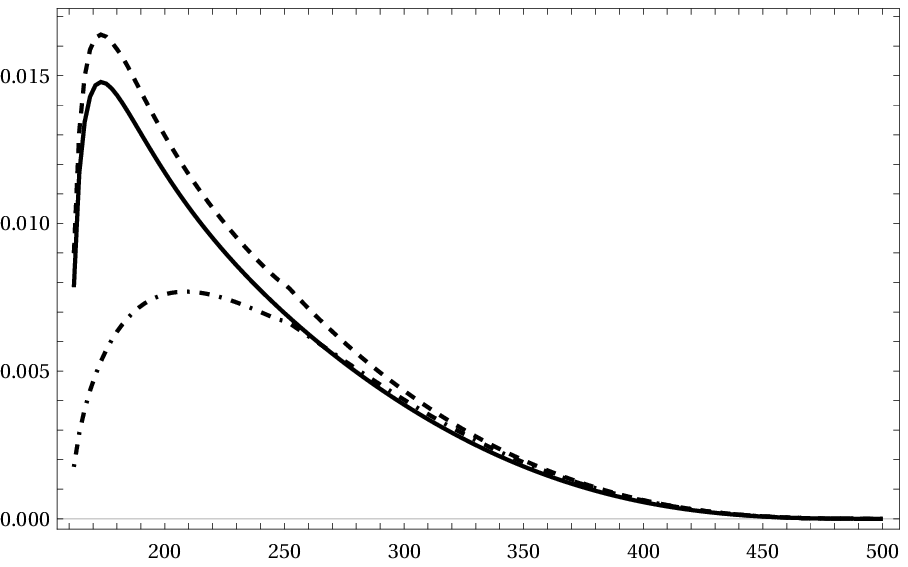}
& 
\includegraphics[width=7cm,height=6cm]
{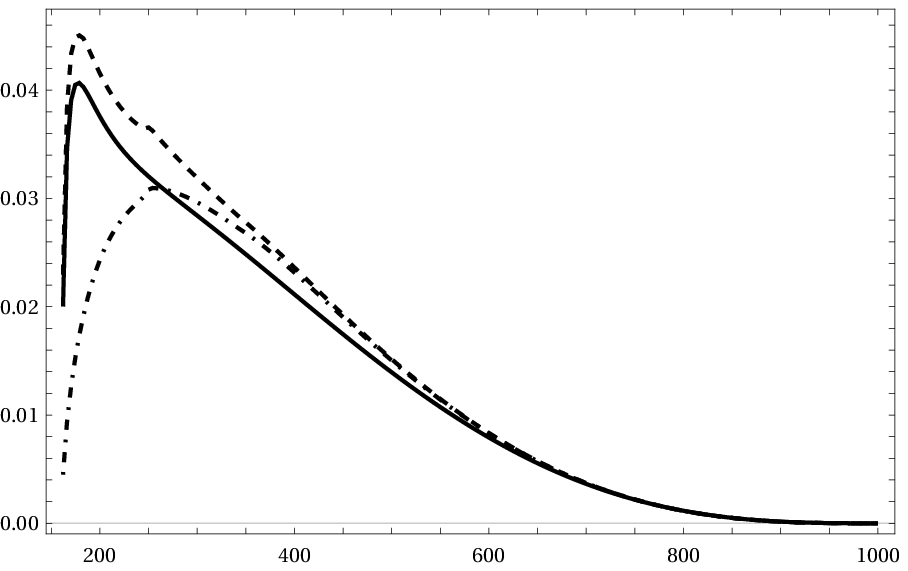}
\\
\hspace{5.2cm} M_{WW} [\text{GeV}]
&
\hspace{5.2cm} M_{WW} [\text{GeV}]
\end{array}$
\caption{\label{decay500} Differential 
cross sections as
a function of $M_{WW}$.}
\end{figure}

\begin{figure}[H]
\centering
$\begin{array}{cc}
\hspace{-3.8cm}
d\sigma/dM_{WW}[\text{fb/GeV}] 
&
\hspace{-3.8cm}
d\sigma/dQ^2[\text{fb/GeV$^2$}] 
\\
\includegraphics[width=7cm,height=6cm]
{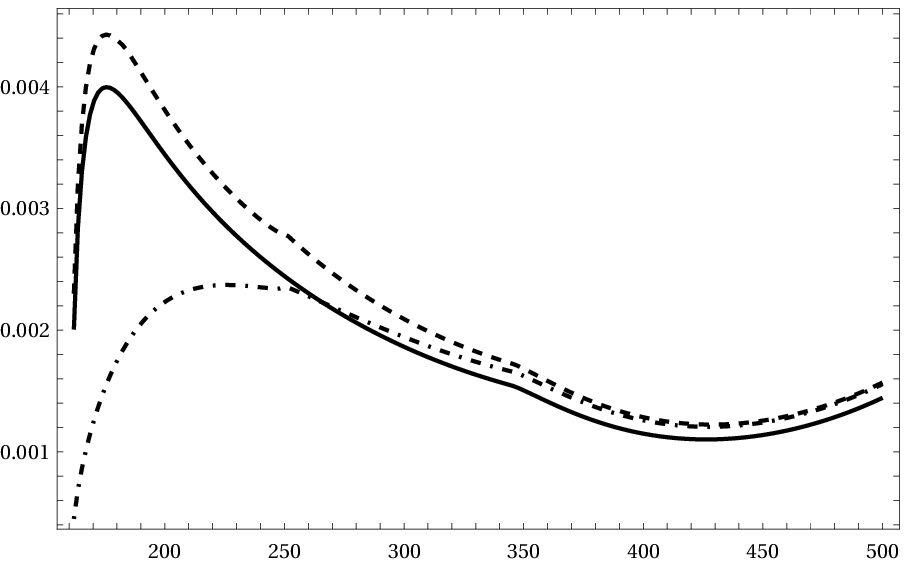}
& 
\includegraphics[width=7cm,height=6cm]
{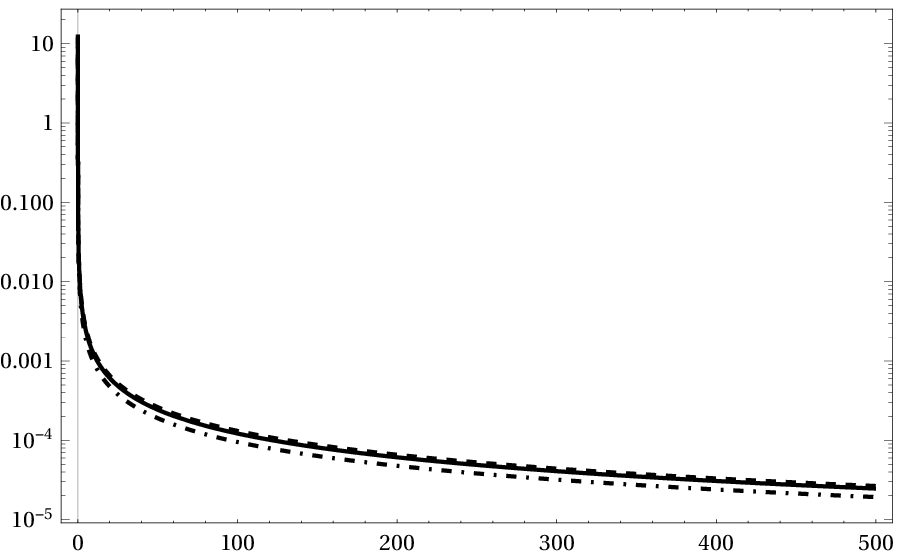}
\\
\hspace{5.2cm} M_{WW} [\text{GeV}]
&
\hspace{5.2cm} Q^2 [\text{GeV}^2]
\end{array}$
\caption{\label{AE_H_EWW} Differential 
cross sections as a function of $M_{WW}$
and $Q^2$.}
\end{figure}

We turn our attention to Higgs 
production at future $e^-$-$\gamma$ colliders.
The signal cross section is written as follows
\begin{eqnarray}
\dfrac{d^2 \sigma (\sqrt{s},Q^2)}
{d M_{WW} \, d Q^2}
&=&
\dfrac{e^2}{16\pi s}
\Bigg[
\dfrac{s^2+(M_{WW}^2 -Q^2-s)^2}
{ Q^2(s^2-Q^2)^2}
\Bigg]\times \Big|F_{00}^{H^*\rightarrow\gamma^*\gamma}
\big(s, Q^2, 0 \big)\Big|^2
\times 
\nonumber\\
&& \times 
\dfrac{2 M_{WW}}
{[(M_{WW}^2-M_H^2)^2 + \Gamma_H^2
M_H^2 ]}\times 
\dfrac{ M_{WW}\; 
\Gamma_{H^*\rightarrow WW}
(M_{WW})}{\pi}. 
\end{eqnarray}
In Fig.~\ref{AE_H_EWW}, we present
differential cross sections with respect
to off-shell Higgs mass $M_{WW}$ 
(left panel) and $Q^2$ (right panel).
In the left Figure, we find same 
previous conclusions. In the right
Figure, cross-section is dominant
in the low region of $Q^2$ in 
comparision
with higher-region of $Q^2$. 
With high-luminosity at future colliders, 
the signal cross sections could be probed 
in the future. The effects of one-loop 
corrections to the off-shell play important 
roles in this analysis for testing the SM 
at higher-energy and  for extracting 
new physic signals. 
\section{Conclusions}   
In this paper, we have presented
full one-loop radiative corrections to 
off-shell decay $H^* \rightarrow W^+W^-$ in 
't Hooft-Veltman within framework of the SM.
One-loop form factors 
for decay process are written in terms 
of the PV-functions in the standard 
notations of {\tt LoopTools}. 
As a result, off-shell decay rates can be 
computed numerically by using this program. 
In phenomenological results, we have shown 
decay rates and one-loop corrections as 
a function of off-shell Higgs mass. 
The corrections are range of 
$5\%$ to $15\%$ for the unpolarized
case for $W$ bosons and of
$-80\%$ to $10\%$ for the longitudinal
polarization
case for $W$ bosons, respectively. 
In applications, we have examined 
the impacts of
one-loop corrections to 
off-shell decay $H^* \rightarrow W^+W^-$
in Higgs processes at future  
colliders.The signal processes such as
$e^-e^+\rightarrow
ZH^*\rightarrow Z(WW)$ and 
$e^-e^+\rightarrow \nu_e\bar{\nu}_e H^*
\rightarrow \nu_e\bar{\nu}_e (WW)$ 
and $e^-\gamma \rightarrow e^-H^*
\rightarrow e^-WW$
are examined. We find that the 
effects are visible impacts
and these should be taken 
into account 
at future colliders. 
\\

\noindent
{\bf Acknowledgment:}~
This research is funded by Vietnam National
University, Ho Chi Minh City (VNU-HCM) 
under grant number C$2022$-$18$-$14$.\\ 
\section{Appendix $A$: Numerical checks}
In order to confirm the analytic 
results presenting in this paper, 
we first  check the $UV$-finiteness of the 
results. It is mentioned in the previous 
section, the form factors $F_{00}^{(G_j)}$
for $j=1,2,3$ contain $UV$-divergent.
To regularize $UV$-divergent, 
counter-term form factor 
$F_{00}^{(G_0)}$ is taken into
account (seen~\ref{counter} for its analytical
formulas).  The numerical results 
for this check are presented in 
the following Table~\ref{UV}. 
In this Table, we change 
$C_{UV}, \mu^2$ and we verify that 
the total form factor $F_{00}$ is very 
good stability (over $11$ digits). 
\begin{table}[H]
\begin{center}
\begin{tabular}{l@{\hspace{2cm}}l}  
\hline \hline 
$(C_{UV}, \mu^2)$
& $\sum\limits_{j=1}^{2}F_{00}^{(G_j)}$\\
& $F_{00}^{(G_0)}$\\
&$F_{00}=\sum\limits_{j=0}^{2}
F_{00}^{(G_j)}$  
\\ \hline \hline
$(0, 1)$ 
& $ -14447.359832765836 
+ 10323.270102389799 \, i $ \\
& $ 14595.587461604524 
+ 0 \, i $ \\
& $ 148.22762883868745 
+ 10323.270102389799 \, i $\\ \hline \\
$(10^2, 10^4)$ 
& $ 100549.21840082797 
+ 10323.270102389799 \, i $ \\
& $ -100400.99077198938 
+ 0 \, i $ \\
& $ 148.22762883868804 
+ 10323.270102389799 \, i $\\ \hline \\
$(10^4, 10^8)$ 
& $ 1.0534774386626098 \times 10^7 
+ 10323.270102389799 \, i $ \\
& $ -1.053462615899727 \times 10^7
+ 0 \, i $ \\
& $ 148.22762883868868 
+ 10323.270102389799 \, i $\\ \hline \\
\end{tabular}
\caption{\label{UV} Checking for 
the UV-finiteness of the 
results at $M_{WW} = 250$ GeV
($p^2=M_{WW}^2$). 
In this case, two real 
bosons are considered in final state.}
\end{center}
\end{table}
\begin{table}[H]
\begin{center}
\begin{tabular}{l@{\hspace{2cm}}l
@{\hspace{2cm}}l@{\hspace{2cm}}l}  
\hline \hline 
$m_{\gamma}$
& $\Gamma_{\textrm{1-loop}}$
& $\Gamma_{\textrm{soft}}$
& $\Gamma_{\textrm{1-loop+soft}}$  
\\ \hline \hline

$10^{-10}$ & $-8.81550898033887$
& $8.963225158327703$ & $0.1477161779888316$
\\ \hline 
$10^{-14}$    & $-12.8788219949896$ &    
$13.02653817297847$ 
& $0.1477161779888316$ 
\\ \hline 
$10^{-18}$ &
$-16.9421350096404$ 
& $17.08985118762925$
& $0.1477161779888511$
\\ \hline 
\end{tabular}
\caption{\label{IR} Checking for the IR-divergent 
cancellation of the decay rates 
$\Gamma_{\textrm{1-loop}}$ and $\Gamma_{\textrm{soft}}$ at $k_C=0.1$, 
$M_{WW}=500$ GeV.}
\end{center}
\end{table}
\begin{table}[H]
\begin{center}
\begin{tabular}{l@{\hspace{2cm}}l
@{\hspace{2cm}}l@{\hspace{2cm}}l}  
\hline \hline 
$k_c$
& $\Gamma_{\textrm{soft}}$
& $\Gamma_{\textrm{hard}}$
& $\Gamma_{\textrm{soft+hard}}$  
\\ \hline \hline
$0.001$
&$6.931568651002313$
&$4.964814985732469$
&$11.89638363673478$
\\ \hline 
$0.005$ 
&$7.641602129869577$
&$4.254791827098171$
&$11.89639395696775$
\\ \hline 
$0.01$
&$7.947396904665006$
&$3.949009959152349$
&$11.89640686381736$
\\
\hline \hline
\end{tabular}
\caption{\label{kc} Checking for the $k_C$-independent of the decay rates $\Gamma_{\textrm{soft}}$ and $\Gamma_{\textrm{hard}}$ at $M_{WW}=500$ GeV, photon mass $m_{\gamma}$ is $10^{-10}$ GeV.}
\end{center}
\end{table}
\section*{Appendix $B$: Counter term}
Counter-term of 
$H \cdot W^+_\mu \cdot W^-_\nu$ 
vertex is taken the form of  
\begin{eqnarray}
\label{counter}
F_{00}^{(G_0)} =
g_{HWW}
\big(
\delta Y
+
\delta G_2
+
\delta G_W
+
2 \delta Z_{W}^{1/2}
+
\delta Z_H^{1/2}
\big).
\end{eqnarray}
\begin{figure}[ht]
\centering
\includegraphics[width=6cm, height=3cm]
{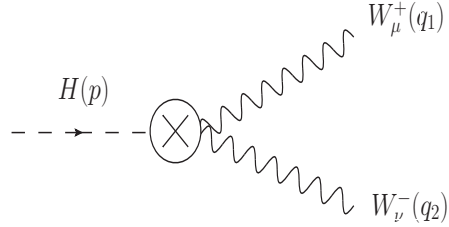}
\caption{Group $0$: counter-term Feynman
diagram.}
\end{figure}
All the renormalization constants presented
in the above formulas can be found in \cite{Tran:2022fdb}. 
\section*{Appendix $C$: Feynman diagrams}
We show all Feynman diagrams for this
decay process in 't Hooft-Veltman gauge
in this appendix. 
\begin{figure}[ht]
\centering
\includegraphics[width=15.0cm, height=3cm]
{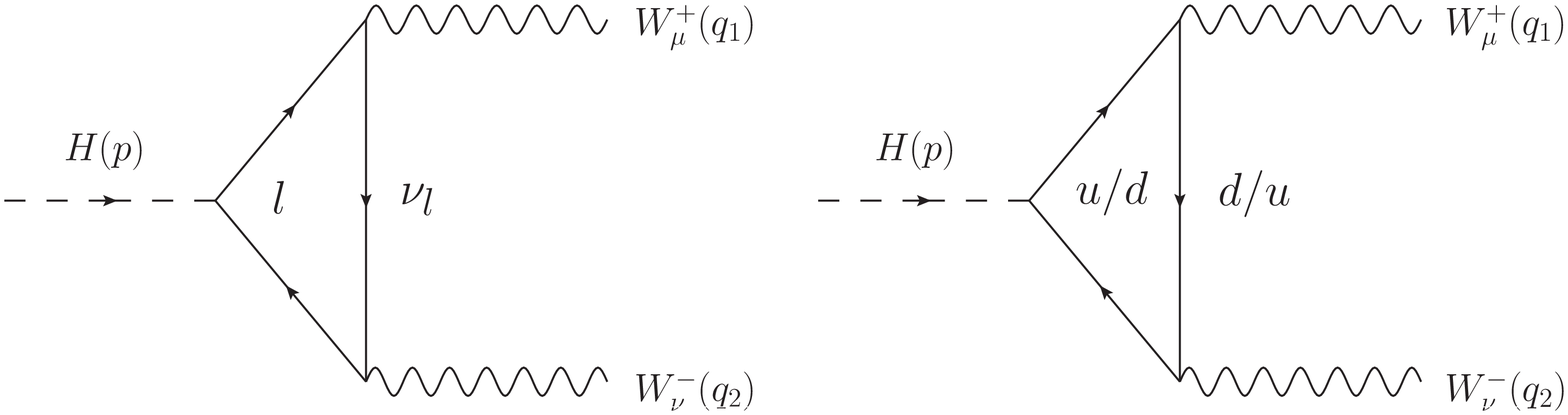}
\caption{Group 1: one-loop Feynman
diagrams with exchanging doublet 
fermions in the loop.}
\end{figure}
\begin{figure}[ht]
\centering
\includegraphics[width=15.0cm, height=2.5cm]
{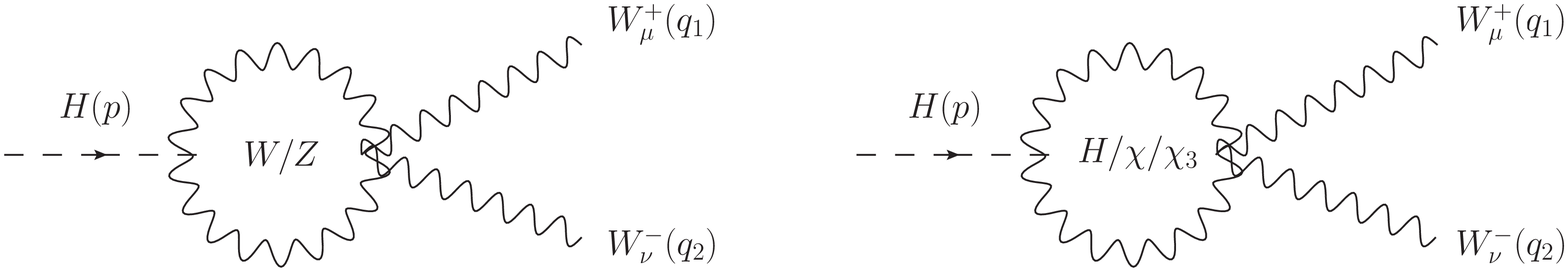}
\\
\includegraphics[width=15.0cm, height=3.cm]
{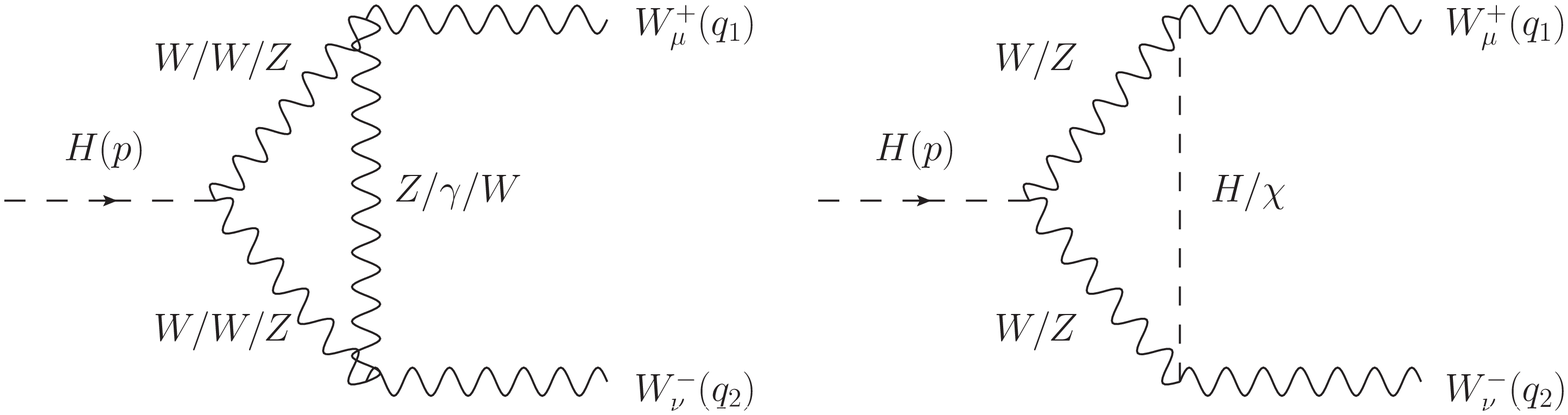}
\\
\includegraphics[width=15.0cm, height=3.cm]
{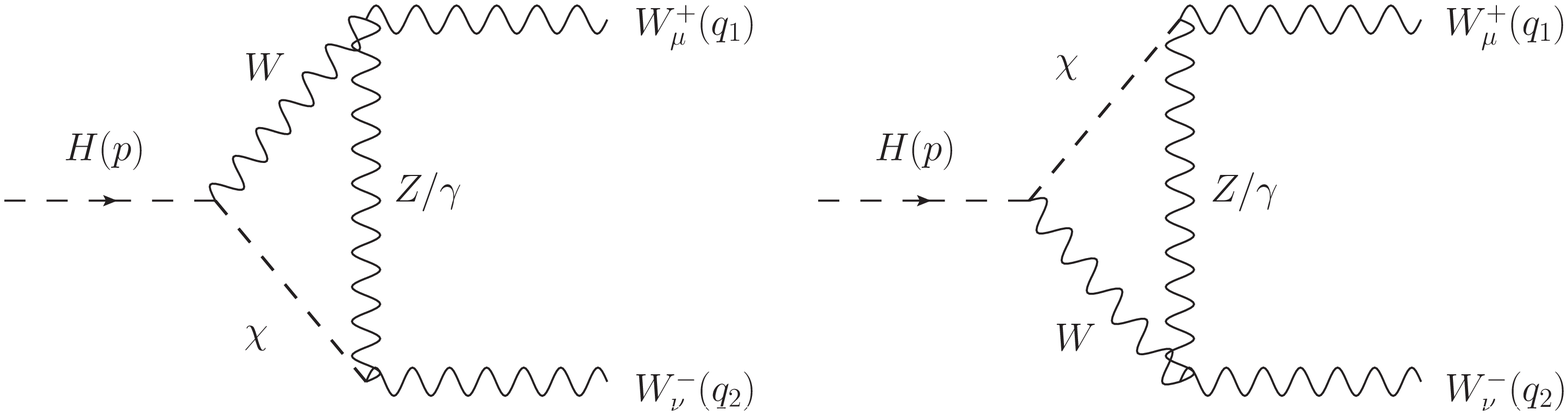}
\\
\includegraphics[width=15.0cm, height=3.cm]
{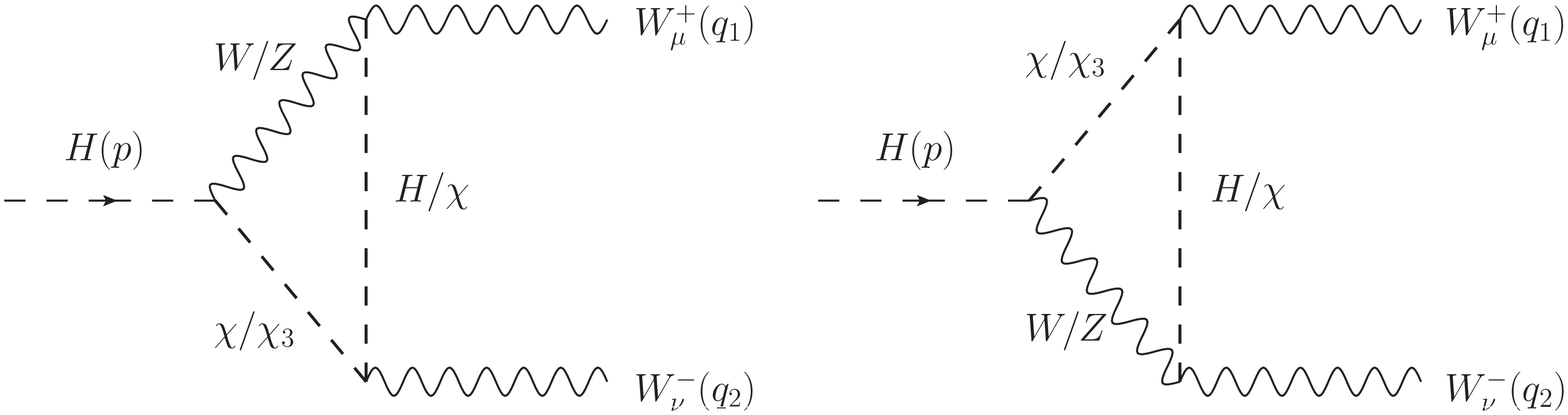}
\\
\includegraphics[width=15.0cm, height=3.cm]
{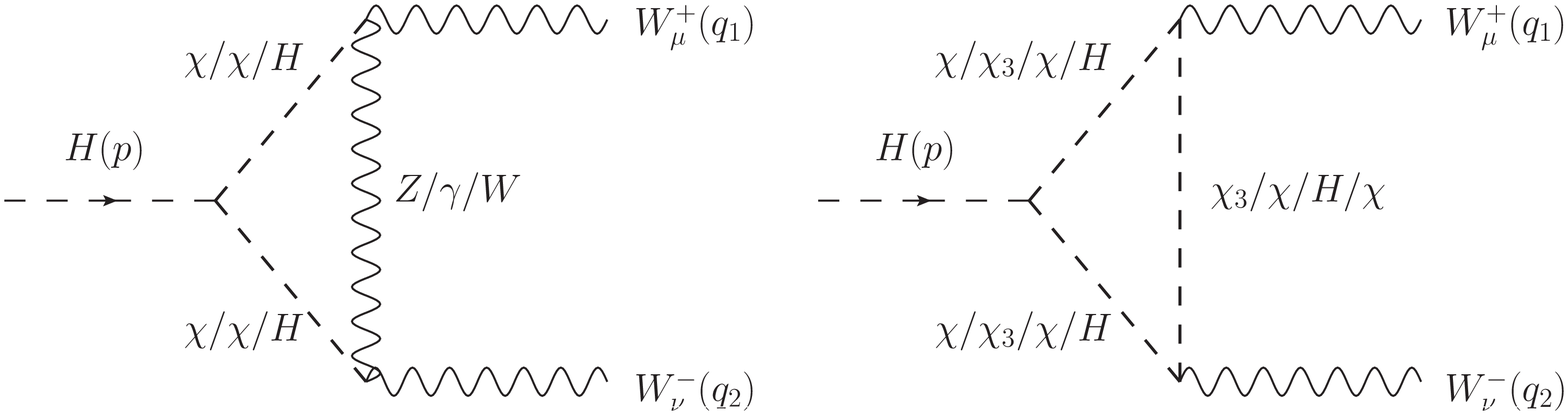}
\\
\includegraphics[width=14.0cm, height=3.cm]
{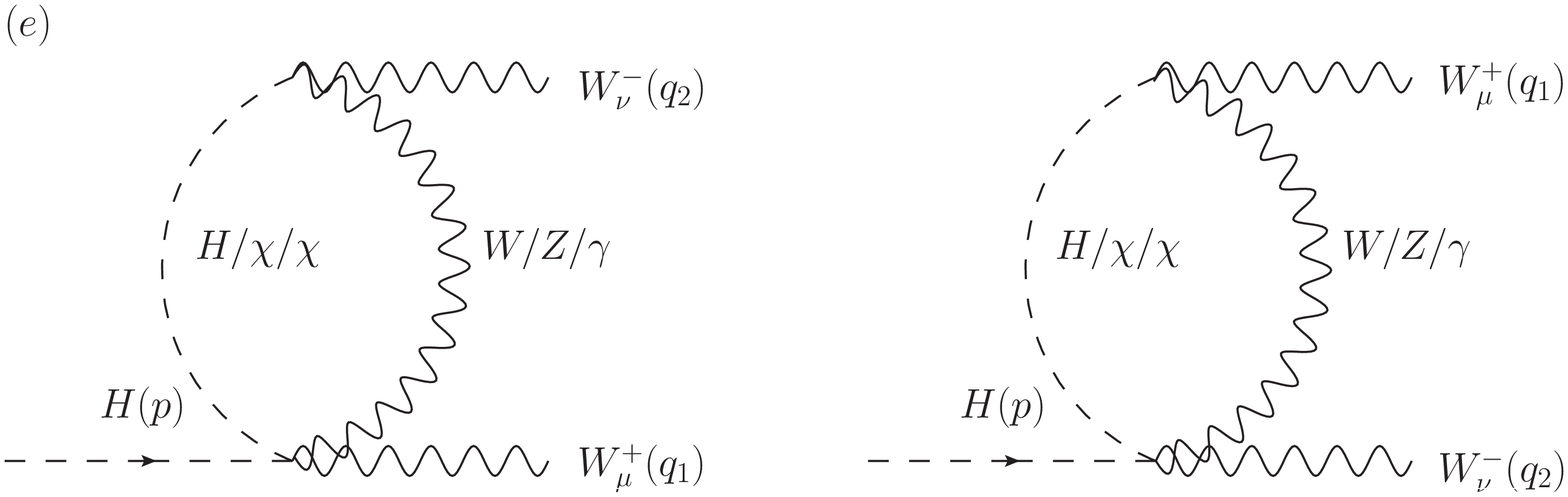}
\\
\includegraphics[width=15.0cm, height=3.cm]
{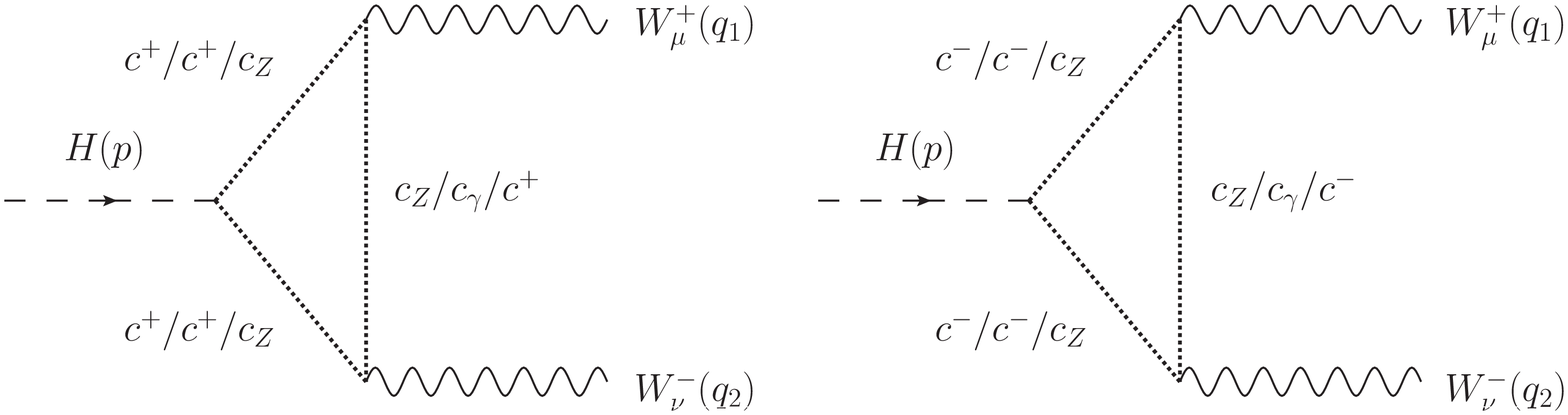}
\caption{Group 2: one-loop Feynman
diagrams with exchanging boson 
and ghost particles
in the loop.}
\end{figure}

\end{document}